# In-situ sputtering from the micromanipulator to enable cryogenic preparation of specimens for atom probe tomography by focused-ion beam

James O. Douglas[1], Michele Conroy[1], F. Giuliani[1], B. Gault[1,2]

[1] Department of Materials, Royal School of Mines, Imperial College London, Prince Consort Road, London SW7 2BP, UK

[2] Max-Planck-Institut für Eisenforschung GmbH, Max-Planck-Str. 1, 40237 Düsseldorf, Germany

## Abstract
Workflows have been developed in the past decade to enable atom probe tomography analysis at cryogenic temperatures. The inability to control the local deposition of the metallic precursor from the gas-injection system (GIS) at cryogenic temperature makes the preparation of site-specific specimens by using lift-out extremely challenging in the focused-ion beam. Schreiber et al. exploited redeposition to weld the lifted-out sample to a support. Here we build on their approach to attach the region-of-interest and additionally strengthen the interface with locally sputtered metal from the micromanipulator. Following standard focused-ion beam annular milling, we demonstrate atom probe analysis of Si in both laser pulsing and voltage mode, with comparable analytical performance as a pre-sharpened microtip coupon. Our welding approach is versatile, as various metals could be used for sputtering, and allows similar flexibility as the GIS in principle.

## 1 Introduction

Atom probe tomography (APT) is a spatially resolved time of flight mass spectrometry analysis technique, where ions are field evaporated from the apex of a nanoscale needle shaped specimen through the application of voltage or laser pulses (Gault et al., 2021). APT analysed volumes are typically of the order of 50 x 50 x 80 nm and are able to give a combination of high compositional (tens of appm(Haley et al., 2020)) and spatial resolution (De Geuser & Gault, 2020; Jenkins et al., 2020), allowing the characterisation of complex nanoscale features in three-dimensions.

The range of materials that APT has been applied to has extended from metals and alloys (Marquis et al., 2013; Ringer, 2006; Blavette et al., 2000) to semiconductor devices (Kelly et al., 2007), insulators (Clark et al., 2016), geological materials (Saxey et al., 2018) and biological materials (Grandfield et al., 2022). The broadening of the field of application has been primarily enabled by the spread of specimen preparation by focused-ion beam (FIB) (Larson et al., 1998; Prosa & Larson, 2017). The combination with a scanning-electron microscope (SEM-FIB) has permitted the precise positioning of a specific, nanoscale region of interest within the apex of the needle shaped specimen and this approach is now considered the routine preparation route for APT (Felfer et al., 2012) and also for transmission electron microscopy (TEM) (Mayer et al., 2007).

This process is typically carried out at room temperature as the commonly used lift-out methodology requires the use of a gas injection system (GIS) projecting an organometallic, precursor gas over the sample (Thompson et al., 2007). The precursor is decomposed by the secondary electrons reemitted by the surface illuminated by an incoming electron or ion beam over specific regions. The decomposition products form a mechanically stable and electrically conducting material (Park et al., 1999). The GIS is used to deposit protective material on defined areas, prior to FIB-milling, to reduce ion beam damage during sample preparation, and also to attach, or weld, micromanipulators to APT-cantilevers or TEM-lamellae for lift-out, and deposit them onto an appropriate support structure. The deposited material has a variable metal content and distribution depending on the precursor gas and the parameters of the incident beam. There has been work on increasing the metal content through precursor gas selection(Diercks et al., 2017) but a significant level of carbon is retained within the deposited material.

This approach has been applied to a wide variety of material systems that are stable at room temperature and ambient environmental conditions. In recent years, advances in instrumentation and increased interest in the analysis of materials that require cryogenic conditions has led to new avenues and associated challenges in nanoscale characterisation.

Cryogenic sample preparation for TEM has revolutionised the field of biological sciences, with the 2017 Nobel prize in Chemistry being awarded to the development of cryo-electron microscopy for the high-resolution structure determination of biomolecules in solution (Anon, 2017). The cryogenic vitrification of hydrated biological materials has facilitated maintaining the samples pristine state, and thus avoiding the dehydration and reducing radiation-induced damage of the sample during analysis (Taylor & Glaeser, 1974, 1976). In the pivotal study by Henderson et al. (Henderson et al., 1990), he emphasised the point that the cryogenic specimen-preparation improvements facilitated the development of cryo-EM to a general technique. To date, the most commonly used cryo-TEM sample preparation technique for biological samples is plunge freezing on TEM grids (Dubochet et al., 1982; Dubochet, 2016). More recently the research field of geochemistry utilised this type of cryogenic TEM sample preparation to answer fundamental physical material science question such as crystal growth and dissolution mechanisms for hydrated materials (Conroy et al., 2017; Kumar et al., 2008; Zhu et al., 2021; Revealed et al., 2010). The use of cryogenic sample preparation for TEM has become "routine' for many research fields but understanding the transfer of experimental observations to real-world interfaces and material performances are shared challenges in both biological and material sciences. Zachman et al showed the first site specific cryogenic FIB sample preparation for TEM of a solid-liquid interfaces of a battery (Zachman et al., 2018).

For APT, it has been shown that cryogenic FIB milling of specimens reduces ion damage and reduces hydrogen ingress during the final stages of specimen sharpening or thinning (Chang et al., 2019), which has been instrumental in allowing the investigation of hydrogen distribution in e.g. Ti- (Chang et al., 2019) or Zr-alloys (Mouton et al., 2021). At lower temperatures, some of the deleterious effects of the chemically-active, most commonly used gallium, for example its diffusion to structural defects in aluminium, can also be avoided (Lilensten & Gault, 2020).

Cryogenic FIB sample preparation was often limited to the final stage of APT sample sharpening or TEM lamellae thinning as performing the full lift-out is challenging. A major challenge with a full cryogenic workflow is that the precursor gas deposits rapidly onto any exposed surface cooled below its condensation temperature (Perea et al., 2017), leading to uncontrolled uniformity and thickness. Although this condensed material can be locally 'cured' to a conducting solid material through the application of an electron beam or ion beam, this requires care to ensure the material is 'cured' for the full thickness and does not contain any residual condensed gas that may expand upon warming

(PARMENTER & NIZAMUDEEN, 2020). This larger scale deposition is extremely quick and has been proposed to reduce time for such areas as electron beam lithography (Salvador-Porroche et al., 2020) but for the purposes of APT sample preparation(Córdoba et al., 2019; Orús et al., 2021), it has limited application due to the lack of site specificity and reduction in thickness control.

Reducing or removing completely the possibility to confine the deposition to a specific area, makes it impossible to weld the region-of-interest to micromanipulator or support and hence to perform site-specific specimen preparation. The spread of Xe-plasma FIBs (PFIB), which can achieve very high ionic currents and hence allow for the removal of large volumes of materials within reasonable times, has offered opportunities to avoid the lift-out and adapt the 'moat approach' (Miller et al., 2005). Halpin et al. introduced this approach(Halpin et al., 2019), later adapted for target grain boundary analysis (Famelton et al., 2021) and for cryogenic temperatures for the analysis of frozen liquids and liquid-solid interfaces (El-Zoka et al., 2020).

The first breakthrough in the preparation of site-specific APT specimens by cryo-lift-out FIB was by Schreiber and co-workers (Schreiber et al., 2018) who used the redeposition from thin FIB cuts across the sample's region-of-interest and the micromanipulator or sample support (e.g. Si-coupon) to weld the parts together. Similar approaches have been used for cryo-specimen preparation by FIB in the biological sciences (Parmenter & Nizamudeen, 2020). Redeposition is caused by the incoming ion beam and causes a mixing of the sputtered materials with the incoming beam (Cairney & Munroe, 2003) and had already been used to create welds for lift-outs (Montoya et al., 2007; Kuba et al., 2020) or fill pores to facilitate further specimen preparation (Zhong et al., 2020). These approaches make use of cryogenically-cooled micromanipulators as well as the cryo-stage, which are not as readily available apart from on dedicated systems. It is also expected that the strength, uniformity, conductivity and overall stability of the weld between the support and the material's region of interest will be highly dependent on the redeposition conditions, i.e. a combination of the primary incoming beam's ions, energy, current and the pattern used, but also the properties of the material being investigated (Winter & Mulders, 2007; Bhavsar et al., 2012).

Here, aiming to facilitate the future establishment of standardised and reproducible fully cryogenic-preparation process flows for APT specimen preparation, we introduce a cryogenic lift-out process. This has the key requirement of a controlled, localised deposition of a mechanically stable and conductive material. We demonstrate a process flow for a fully cryogenic, GIS-free lift-out, mounting and sharpening of viable atom probe samples with sufficient mechanical stability and electrical conductivity to be analysed by both laser and voltage pulsing.

## 2 Materials & instruments

A Helios Hydra CX (5CX) plasma FIB from Thermofisher Scientific with an Aquillos cryo stage equipped with an Easy Lift tungsten cryo micromanipulator, was used for sample prepration in this work. The FIB column is set at 52 degrees to the electron column. Xe plasma was used throughout this process.

A commercially available highly Sb doped single crystal silicon micropost array (Cameca) was used as both the substrate and for mounting APT samples for this demonstration, as these arrays are used as standard reference materials for LEAP analysis for both voltage and laser pulsing due to their mechanical stability and high levels of conductivity. This also removes the requirement for swapping the substrate for the sample mount during the cryo lift-out process, however this would not be a limitation for the proposed approach. The array was loaded into a Cu clip mount (Cameca) and placed into a specimen holder, i.e. puck, which was then inserted into the Aquillos stage. The stage and micromanipulator were then cooled to approximately 90 K by using a circulation of gaseous nitrogen passing through a heat exchanged system within a liquid nitrogen (LN$_2$) dewar. A N$_2$ gas flow of 180 mg/s was used to achieve this base temperature. Atom probe analysis was performed on a Cameca Local Electrode Atom Probe 5000XR. Both laser pulsing (30 pJ, 140–200 kHz, 1 ion per 100 pulses on average, 50 K base temperature) and voltage pulsing (20% pulse fraction, 200 kHz, 1 ion per 100 pulses on average, 50 K base temperature) have been tried.

# 3 Methods

## 3.1 Cryo-lift-out

Flat top silicon posts were prepared for lift-out by milling them at zero degrees stage tilt with a 30 kV 4 nA probe, corresponding to a 52˚ angle with respect to the specimen's normal, such that the apex diameter is approximately 5–6 µm and with an angled surface. This preliminary step was proposed by Schreiber (Schreiber et al., 2018) to maximise the surface contact between the lifted-out wedge and the post to strengthen the weld. This step can be performed at room temperature prior to cooling the stage.

All further steps were done with the stage and micromanipulator at the lowest temperature setting (approximately 95K). The undercut and lift-out procedure were carried out at zero degrees stage tilt in order to match the angle of the wedge with the prepared silicon post. A cantilever of 30 µm x 5 µm x 10 µm was prepared using milling patterns typically used for APT sample preparation, simply larger than typical (Thompson et al., 2007), Figure 1 (a) and with sufficiently large trenches to be able observe the bottom of the cantilever during the undercut stage to ensure sample release. The manipulator was prepared through sharpening to a point less than 5µm in diameter and placed in direct contact with a flat side parallel to the side of the cantilever to maximise the contact area for the subsequent redeposition. A series of 6–8 line or small and thin rectangles patterns were then milled across the interface between the manipulator and the cantilever such that a small amount of material is sputtered between them using a 30 kV 100 pA probe, forming a weld. Only a small amount of material is required to make a connection that is mechanically sufficiently stable for the lift-out process and the exact probe current and size/shape of lines may vary with the relative sputtering rates of the two materials to be joined. This weld is expected to still be significantly less stable than a standard GIS glue section due to the lack of a continuous solid connection between the support structure and the sample material. Once welded, the arm of the cantilever is finally cut, and the wedge is lifted out, Figure 1 (a).

The wedge is then placed directly onto the prepared silicon post, Figure 1 (b), with care taken to align and make contact with the post in a single motion, as moving making contact and moving away from the post can cause the cantilever to pivot or break away due to e.g. local Van der Waals forces or electrostatics from possible charging. Due to the fragile nature of this connection, any motion to the manipulator caused by the changing of FIB probe current aperture can be sufficient to cause sample loss and so it is recommended to not change probe current if the specific changes cause vibration. Full thermalisation of the stage carrying the sample holder and support and the manipulator should also be established prior to lift-out to avoid any degree of mechanical drift during that could also stress the weld.

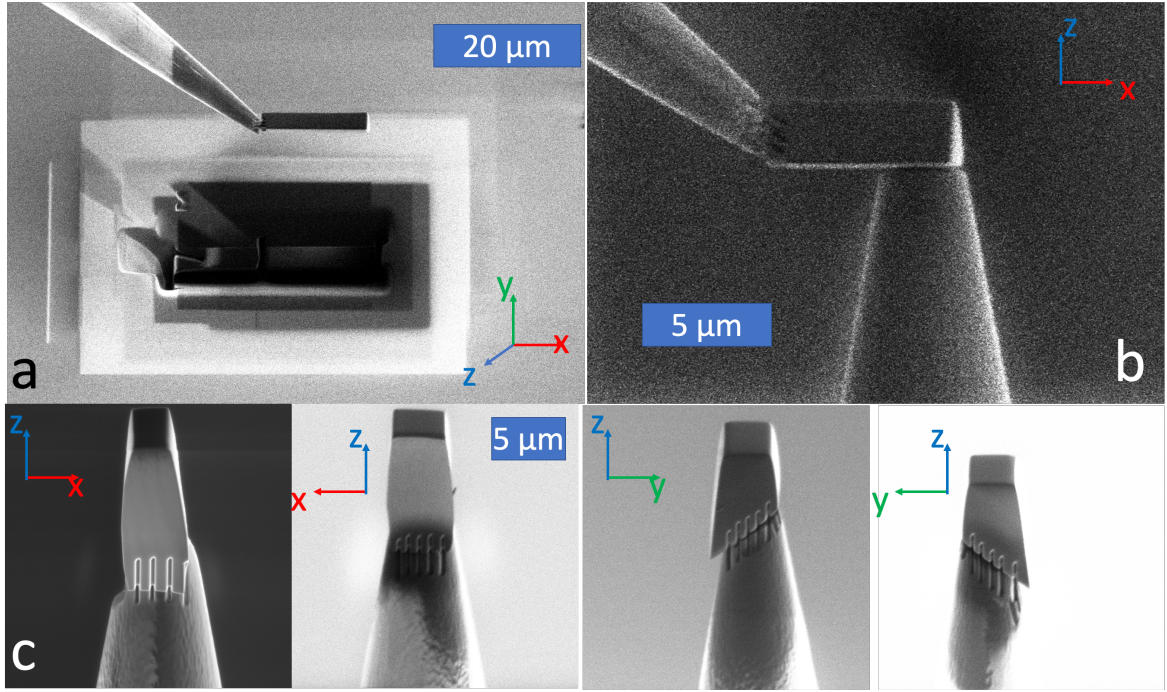

*Figure 1: (a) cantilever containing the region-of-interest lifted out from the sample. The cantilever was attached by using redeposition. (b) cantilever positioned onto the pre-shaped post of a commercial microtip coupon ready for attachment. (c) view of the weld by redeposition from the four sides performed in succession.*

Once in place, and depending on the size of the wedge, a similar series of 4–8 lines (30 kV 0.1 nA for a line feature with a depth of 0.1 μm) were used to repeat the redeposition process on each of the four sides of the lifted-out wedge connecting to the post, Figure 1 (c). Care must be taken to only cause sufficient milling to connect the structures and not mill through the post. This redeposition stage is only to maintain stability of the interface for the next stage. Experimentation showed that although the edges of wedges mounted in this way made sufficient connection to maintain stability, there can be significant gaps between the wedge and the post in the centre. This reduces contact area, which will reduce mechanical stability and electrical conductivity for subsequent necessary for atom probe analysis.

Once all four sides of the wedge have been connected to the post using redeposition, a trapezium-shaped region is milled from the side of the wedge such that the region-of-interest at the top of the wedge is retained. The manipulator is then inserted into this gap, Figure 2(a), such that it is almost in contact with the wedge with a distance of a few hundred of nanometres. A cross sectional, single pass mill pattern at 30 kV 1 nA is then applied to the edge of the manipulator, with the ion beam rastering towards (pink arrow) the manipulator in order to maximise the sputtering from the manipulator and minimise the milling from the recently redeposited material. The precise conditions will depend on the currents available and the focusing ability of the probe current used on the instrument. The manipulator is then moved closer to maintain the distance between itself and the milled-out region and the milling process is repeated. The deposited material can be seen to slowly fill in the gap within the milled-out region and to have generally much higher contrast than the surrounding silicon, with some contrast variations within it which imply some form of microstructure formed during deposition. For a manipulator with an end diameter of approximately 6 μm used herein, sufficient material was deposited within ten minutes. For comparison, deposition of a 2 μm x 2 μm x 0.5 μm Pt using 12 kV 30 pA $Xe^+$ ions on either side of a lift-out standard wedge deposited on a commercial coupon requires approximately 6 minutes each side, and so the total time for this in situ deposition is similar.

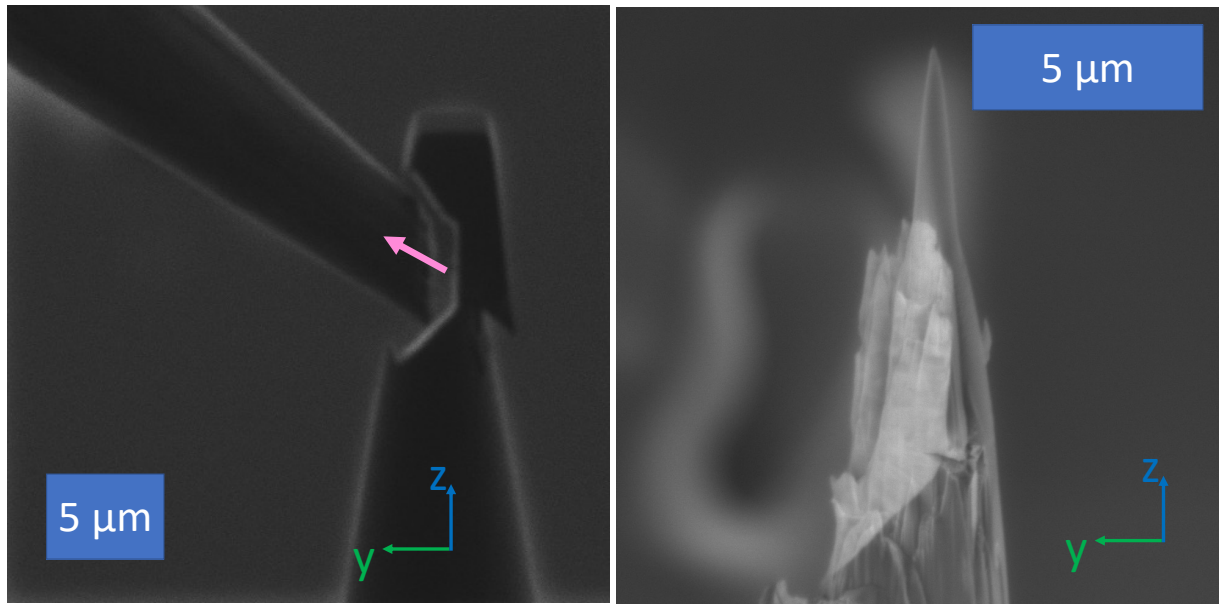

Figure 2: (a) insertion of the micromanipulator in the milled gap to enable the in situ strengthening weld. (b) final specimen following needle-shaping by annular milling; the W-weld is imaged brightly owing to the back-scattered contrast.

Once sufficient redeposited material was in place to fill in the milled-out region, the sample was sharpened to an apex diameter less than 100 nm using 30 kV $Xe^+$ from 1 nA to 30 pA and polishing using 5 kV $Xe^+$ ions at 30 pA to remove regions severely damaged by the incoming energetic ion beam. Care was taken to ensure that minimal redeposited material was removed from the interface section to allow the largest possible volume of material for thermal and electrical conductivity.

### 3.2  Transfer and atom probe analysis

The sample was then allowed to warm up under vacuum within the chamber to room temperature by removing the heat exchange unit from the LN2 dewar while continuing a $N_2$ gas flow of 120 mg/s before being transferred via a FerroVac vacuum transfer system to the loadlock chamber of the LEAP 5000XR. Transfer at cryogenic temperature was avoided for this demonstration in order to remove the possibility of adsorbed gas species and other material onto these demonstration specimens during transport of a cooled sample through non cooled regions of the transfer route within the FIB and LEAP systems. The sample was transferred into the analysis chamber of the atom probe and data collected using laser mode pulsing and then voltage pulsing in order to compare data quality from both approaches with standard silicon pre-sharpened micro tips (Cameca).

The initial laser analysis using commonly used conditions showed comparable data quality in terms of peak shape to that collected by commercial silicon pre-sharpened microtips, Figure 3 (a–b). This indicates that the overall thermal conductivity of the sample, including the redeposited tungsten material, is sufficiently high to not cause visible delayed evaporation events, also known as "thermal tails", in the recorded data. The subsequent voltage pulsing analysis, carried out up to a standing voltage of 9 kV with a pulse fraction of 20 %, also showed comparable quality data to that collected from pre-sharpened microtips, Figure 3 (c–d). The sample survived analysis at voltages at the upper range of the requirements for most materials (approximately 11kV), which can be taken as an indication that the mechanical stability of the sample is sufficient for the stresses associated with those voltages.

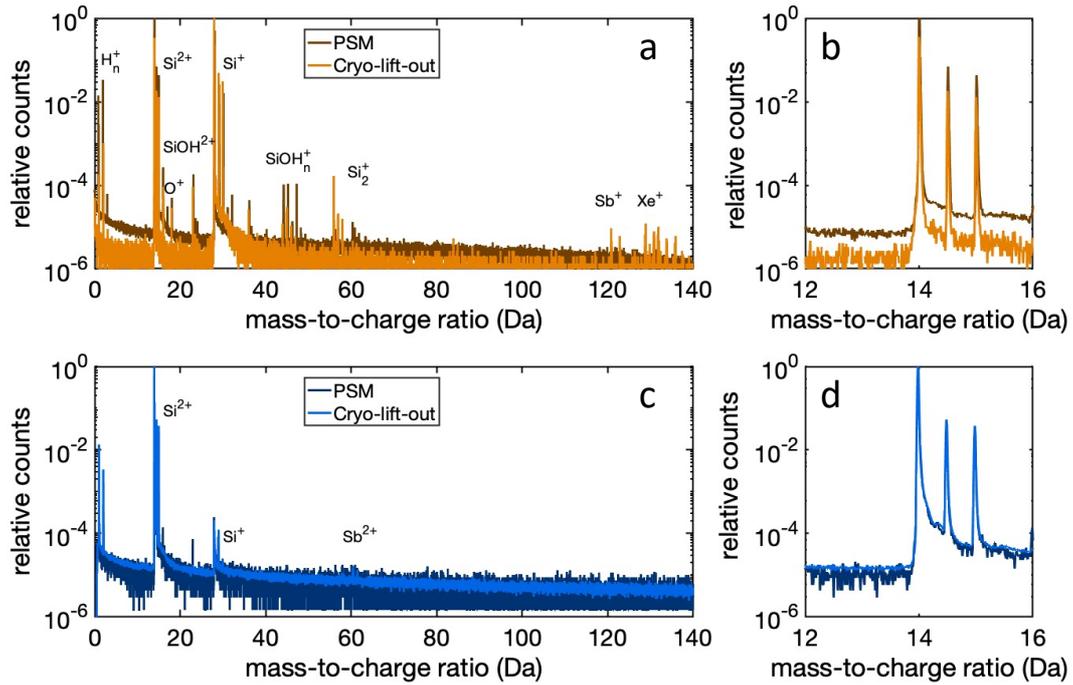

*Figure 3: normalised mass spectra obtained from the analyses of a pre-sharpened microtip (PSM) and specimens prepared by cryo-lift-out from the Sb-doped Si: (a) in laser pulsing mode with (b) a close-up on the $Si^{2+}$ peaks; (c) in high-voltage pulsing mode with (d) a close-up on the $Si^{2+}$ peaks.*

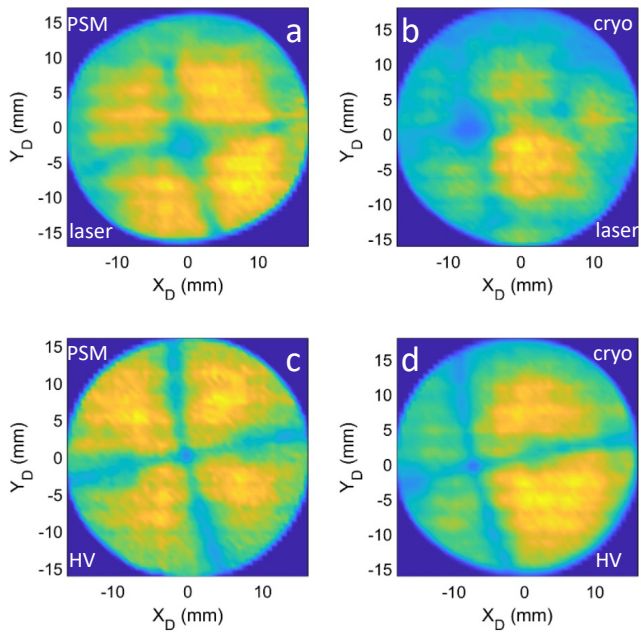

*Figure 4: detector hit histogram from the analyses in laser pulsing mode for (a) a pre-sharpened microtip and (b) a specimen prepared by cryo-lift-out; and in voltage pulsing mode for (c) a pre-sharpened microtip and (d) a specimen prepared by cryo-lift-out.*

Finally in Figure 4 are plotted the detector hit maps in laser (a–b) and HV mode (c–d) for both pre-sharpened microtips and specimens prepared by cryo-lift-out. Since specimens were prepared from a [001]-oriented wafer, the 4-fold symmetry of the (004) set of planes is readily visible in all maps.

The position of the main pole in the map reveals the relative orientation of the lifted-out specimen with respect to the main specimen axis (z in Figure 1 and Figure 2) (Gault et al., 2012). We demonstrate here a reproducible and consistent positioning of lifted-out wedge with respect to the support. The difference in apparent the angular field-of-view and sharpness of the pole figure is simply related to the specimen's radius and analysis conditions (i.e. laser vs. HV) as reported previously (Gault et al., 2010).

# 4   Discussion

Localised redeposition has been shown to be viable in the lift-out and mounting of TEM lamellae onto grids (Kuba et al., 2020). This is where two features, typically a manipulator and lamella or a lamella and a grid arm, are placed in close proximity and an ion beam is used to remove small amounts of material from each side such that redeposits on and in-between the two features, forming a sufficiently stable bridge, i.e a weld with sufficient mechanical stability to allow for further lamella thinning and TEM analysis. If routine for TEM – it is part of the routine training procedure by one of the leading brand of focused-ion beams – including at cryogenic temperatures, this process had not been optimised for APT specimen preparation. The requirements for APT specimens are rather different to TEM, and good electrical and thermal conductivity are required to ensure adequate analytical performance, along with sufficient strength to resists the extreme stresses during analysis (Moy et al., 2011; Kölling & Vandervorst, 2009; Wilkes et al., 1972).

The use of the tungsten micromanipulator as a localised sputtering source allows the use of a highly conductive and mechanical stable material for redeposition. It offers the major advantage that this can be performed at cryogenic temperature, and to lead to good adhesion, without the use of C-containing precursors or water, which can both be sources of contamination for some materials. In addition, the material that would normally be redeposited would be provided by the sample itself and in the case of frozen, hydrated samples, this could have significantly reduced electrical conductivity (Schreiber et al., 2018). In-situ electrochemical charging and subsequent cryogenic sample transfer of atom probe samples prepared by lift-out has been shown to have issues with yield loss due to the room temperature GIS deposited weld and the replacement with redeposited pure metal could reduce sample failure (Khanchandani et al., 2022). Avoiding the requirement for a GIS for the lift-out protocol can be generalised for dual-beam instruments that have a micromanipulator but either no GIS or no functioning GIS.

The use of a precursor gas for local deposition is also a potential major source of carbon that can contaminate the surface, which can become critical in the correlative workflows where TEM and APT are combined on the same specimen (Herbig, 2018; Herbig et al., 2015; Sasaki et al., 2022; Spurgeon et al., 2018), that require careful cleaning as discussed extensively by Herbig and Kumar (Herbig & Kumar, 2021). A GIS-free approach can hence be highly advantageous.

In the future, we expect to explore the use of this in situ deposition approach to perform coatings in pre-sharpened specimens, which has been shown to be important for mechanical stability (Seol et al., 2016) and to enable shielding that facilitate analysis of e.g. Li-containing materials (Kim et al., 2022). At this stage, however, we do not know what is the composition or microstructure (e.g grain size & orientation) of the deposited material that forms the weld itself. It would be interesting to find out if Xe is being deposited as well, if it forms gaseous bubbles for instance, that might affect the mechanical stability of the weld, and, finally, how the deposition conditions can be adjusted to control the microstructure of the weld itself.

If tungsten is a standard material for micromanipulators due to their ease of manufacture using electrochemical polishing and resistance to ion milling and have a relatively low cost for replacement. However, manipulators could be manufactured of other metals that would be more suitable for redeposition or pre-fabricate (Kölling & Vandervorst, 2009). Multiple metals have been shown to be suitable for cryogenic sputter deposition (Chambers et al., 1988, 1998), and these will be trialed in the future.

Due to the increase in time required for sample preparation and the increase in width of the lift-out wedge to match the dimensions of the prepared support structure, it is unlikely that the full length of the lifted-out wedge would be usable. So it is likely that a reduction in the length of wedge can be carried out without reducing sample throughput. Preliminary steps such as the preparation of the micromanipulator or of the silicon post are not required to be carried out at cryogenic conditions, having a range of pre-prepared structures can be organised in advance, maybe even in a fully automated process. Prior to attachment, our approach only currently consists of a single, angled cut and there is scope for more elaborate pre-fabricated structures, such as an angled V cut or dovetail, that may be more suitable depending on the requirements for the sample and maybe its tendency for redepositing (Miller et al., 2005).

# 5 Conclusion

To conclude, we have demonstrated a reproducible methodology for the GIS-free, lift-out and mounting of APT specimens at cryogenic temperatures, via selective redeposition of material from a micromanipulator using commercially available instrumentation. This process has been showcased in the analysis of Sb-doped silicon specimens, with thermal and mechanical stability that compares with silicon reference material from pre-sharpened microtips. This approach can be readily adapted for use in a variety of cryogenic sample preparation of APT where existing GIS based methods are not appropriate.

**Acknowledgements**

FG and BG are grateful for funding from the EPSRC under the grant # EP/V007661/1. BG is grateful for funding from the ERC for the project SHINE (ERC-CoG) #771602. MC acknowledges funding from the Royal Society Tata University Research Fellowship (URF\R1\201318) and the EPSRC NAME Programme Grant EP/V001914/1. Dr Guillaume Amiard is acknowledged for fruitful discussions.

**References**

BHAVSAR, S. N., ARAVINDAN, S. & RAO, P. V. (2012). Experimental investigation of redeposition during focused ion beam milling of high speed steel. *Precision Engineering* **36**, 408–413.

BLAVETTE, D., CADEL, E. & DECONIHOUT, B. (2000). The Role of the Atom Probe in the Study of Nickel-Based Superalloys. *Materials Characterization* **44**, 133–157.

CAIRNEY, J. M. & MUNROE, P. R. (2003). Redeposition effects in transmission electron microscope specimens of FeAl–WC composites prepared using a focused ion beam. *Micron* **34**, 97–107.

CHAMBERS, D. L., WAN, C. T., SUSI, G. T. & TAYLOR, K. A. (1988). Sputtering deposition of thin films at cryogenic temperatures. *Surface and Coatings Technology* **36**, 893–900.

———— (1998). Sputter deposition of aluminum and other alloys at cryogenic temperatures. *Journal of Vacuum Science & Technology A: Vacuum, Surfaces, and Films* **7**, 1305.

CHANG, Y., LU, W., GUÉNOLÉ, J., STEPHENSON, L. T., SZCZPANIAK, A., KONTIS, P., ACKERMAN, A. K., DEAR, F., MOUTON, I., ZHONG, X., RAABE, D., GAULT, B., ZHANG, S., DYE, D., LIEBSCHER, C. H., PONGE, D., KORTE-KERZE, S., RAABE, D. & GAULT, B. (2019). Ti and its alloys as examples of cryogenic focused ion beam milling of environmentally-sensitive materials. *Nature Communications* **10**, 942.

CLARK, D. R., ZHU, H., DIERCKS, D. R., RICOTE, S., KEE, R. J., ALMANSOORI, A., GORMAN, B. P. & O'HAYRE, R. P. (2016). Probing Grain-Boundary Chemistry and Electronic Structure in Proton-Conducting Oxides by Atom Probe Tomography. *Nano Letters* **16**, 6924–6930.

CONROY, M., SOLTIS, J. A., WITTMAN, R. S., SMITH, F. N., CHATTERJEE, S., ZHANG, X., ILTON, E. S. & BUCK, E. C. (2017). Importance of interlayer H bonding structure to the stability of layered minerals. *Scientific Reports 2017 7:1* **7**, 1–10. https://www.nature.com/articles/s41598-017-13452-7 (Accessed November 10, 2022).

CÓRDOBA, R., ORÚS, P., STROHAUER, S., TORRES, T. E. & DE TERESA, J. M. (2019). Ultra-fast direct growth of metallic micro- and nano-structures by focused ion beam irradiation. *Scientific Reports 2019 9:1* **9**, 1–10.

DIERCKS, D. R., GORMAN, B. P. & MULDERS, J. J. L. (2017). Electron Beam-Induced Deposition for Atom Probe Tomography Specimen Capping Layers. *Microscopy and Microanalysis* **23**, 321–328. https://www.cambridge.org/core/journals/microscopy-and-microanalysis/article/abs/electron-beaminduced-deposition-for-atom-probe-tomography-specimen-capping-layers/89739766CB3D7E67D29E444354319123 (Accessed September 13, 2021).

DUBOCHET, J. (2016). A Reminiscence about Early Times of Vitreous Water in Electron Cryomicroscopy. *Biophysical Journal* **110**, 756–757.

DUBOCHET, J., LEPAULT, J., FREEMAN, R., BERRIMAN, J. A. & HOMO, J. -C (1982). Electron microscopy of frozen water and aqueous solutions. *Journal of Microscopy* **128**, 219–237.

EL-ZOKA, A. A., KIM, S.-H., DEVILLE, S., NEWMAN, R. C., STEPHENSON, L. T. & GAULT, B. (2020). Enabling near-atomic-scale analysis of frozen water. *Science Advances* **under revi**.

FAMELTON, J. R., HUGHES, G. M., WILLIAMS, C. A., BARBATTI, C., MOODY, M. P. & BAGOT, P. A. J. (2021). Xenon plasma focussed ion beam preparation of an Al-6XXX alloy sample for atom probe tomography including analysis of an α-Al(Fe,Mn)Si dispersoid. *Materials Characterization* **178**, 111194.

FELFER, P. J., ALAM, T., RINGER, S. P. & CAIRNEY, J. M. (2012). A reproducible method for damage-free site-specific preparation of atom probe tips from interfaces. *Microscopy Research and Technique* **75**, 484–491.

GAULT, B., CHIARAMONTI, A., COJOCARU-MIRÉDIN, O., STENDER, P., DUBOSQ, R., FREYSOLDT, C., MAKINENI, S. K., LI, T., MOODY, M. & CAIRNEY, J. M. (2021). Atom Probe Tomography. *Nature Reviews, Methods Primers* 1–51.

GAULT, B., LA FONTAINE, A., MOODY, M. P. P. M. P., RINGER, S. P. P. S. P. & MARQUIS, E. A. A. E. A. (2010). Impact of laser pulsing on the reconstruction in an atom probe tomography. *Ultramicroscopy* **110**, 1215–1222.

GAULT, B., MOODY, M. P., CAIRNEY, J. M. & RINGER, S. P. (2012). Atom probe crystallography. *Materials Today* **15**, 378–386.

DE GEUSER, F. & GAULT, B. (2020). Metrology of small particles and solute clusters by atom probe tomography. *Acta Materialia* **188**, 406–415.

GRANDFIELD, K., MICHELETTI, C., DEERING, J., ARCURI, G., TANG, T. & LANGELIER, B. (2022). Atom probe tomography for biomaterials and biomineralization. *Acta Biomaterialia* **148**, 44–60.

HALEY, D., LONDON, A. J. & MOODY, M. P. (2020). Processing APT Spectral Backgrounds for Improved Quantification.

HALPIN, J. E., WEBSTER, R. W. H., GARDNER, H., MOODY, M. P., BAGOT, P. A. J. & MACLAREN, D. A. (2019). An in-situ approach for preparing atom probe tomography specimens by xenon plasma-focussed ion beam. *Ultramicroscopy*.

HENDERSON, R., BALDWIN, J. M., CESKA, T. A., ZEMLIN, F., BECKMANN, E. & DOWNING, K. H. (1990). Model for the structure of bacteriorhodopsin based on high-resolution electron cryo-microscopy. *Journal of Molecular Biology* **213**, 899–929.

HERBIG, M. (2018). Spatially correlated electron microscopy and atom probe tomography: Current possibilities and future perspectives. *Scripta Materialia* **148**, 98–105. https://www.sciencedirect.com/science/article/pii/S1359646217301422.

HERBIG, M., CHOI, P. & RAABE, D. (2015). Combining structural and chemical information at the nanometer scale by correlative transmission electron microscopy and atom probe tomography. *Ultramicroscopy* **153**, 32–39. http://www.scopus.com/inward/record.url?eid=2-s2.0-84923367338&partnerID=tZOtx3y1.

HERBIG, M. & KUMAR, A. (2021). Removal of hydrocarbon contamination and oxide films from atom probe specimens. *Microscopy Research and Technique* **84**, 291–297. https://onlinelibrary.wiley.com/doi/full/10.1002/jemt.23587 (Accessed November 10, 2022).

JENKINS, B. M., DANOIX, F., GOUNÉ, M., BAGOT, P. A. J., PENG, Z., MOODY, M. P. & GAULT, B. (2020). Reflections on the Analysis of Interfaces and Grain Boundaries by Atom Probe Tomography. *Microscopy and Microanalysis* **26**, 247–257.

KELLY, T. F., LARSON, D. J., THOMPSON, K., ALVIS, R. L., BUNTON, J. H., OLSON, J. D. & GORMAN, B. P. (2007). Atom Probe Tomography of Electronic Materials. *Annual Review of Materials Research* **37**, 681–727.

KHANCHANDANI, H., KIM, S.-H., VARANASI, R. S., PRITHIV, T., STEPHENSON, L. T. & GAULT, B. (2022). Hydrogen and deuterium charging of lifted-out specimens for atom probe tomography. *Open Research Europe* **1**, 122.

KIM, S., ANTONOV, S, ZHOU, X, STEPHENSON, L., JUNG, C, EL-ZOKA, A., SCHREIBER, D K, CONROY, S., GAULT, B, MATER, J., KIM, S.-H., ANTONOV, STOICHKO, ZHOU, XUYANG, STEPHENSON, L. T., JUNG, CHANWON, EL-ZOKA, A. A., SCHREIBER, DANIEL K, CONROY, M. & GAULT, BAPTISTE (2022). Atom probe analysis of electrode materials for Li-ion batteries: challenges and ways forward. *Journal of Materials Chemistry A* **6**, 4883–5230.

KÖLLING, S. & VANDERVORST, W. (2009). Failure mechanisms of silicon-based atom-probe tips. *Ultramicroscopy* **109**, 486–491.

KUBA, J., MITCHELS, J., HOVORKA, M., ERDMANN, P., BERKA, L., KIRMSE, R., KÖNIG, J., DE BOCK, J., GOETZE, B. & RIGORT, A. (2020). Advanced cryo-tomography workflow developments – correlative microscopy, milling automation and cryo-lift-out. *Journal of Microscopy*.

KUMAR, S., WANG, Z., LEE PERM, R. & TSAPATSIS, M. (2008). A structural resolution cryo-TEM study of the early stages of MFI growth. *Journal of the American Chemical Society* **130**, 17284–17286. https://pubs.acs.org/doi/full/10.1021/ja8063167 (Accessed November 10, 2022).

LARSON, D. J., FOORD, D. T., PETFORD-LONG, A. K., ANTHONY, T. C., ROZDILSKY, I. M., CEREZO, A. & SMITH, G. W. D. (1998). Focused ion-beam milling for field-ion specimen preparation: *Ultramicroscopy* **75**, 147–159.

LILENSTEN, L. & GAULT, B. (2020). New approach for FIB-preparation of atom probe specimens for aluminum alloys. *PLoS ONE* **15**.

MARQUIS, E. A., BACHHAV, M., CHEN, Y., DONG, Y., GORDON, L. M. & MCFARLAND, A. (2013). On the current role of atom probe tomography in materials characterization and materials science. *Current Opinion in Solid State and Materials Science* **17**, 217–223.

MAYER, J., GIANNUZZI, L. A., KAMINO, T. & MICHAEL, J. (2007). TEM Sample Preparation and Damage. *MRS Bulletin* **32**, 400–407.

MILLER, M. K., RUSSELL, K. F. & THOMPSON, G. B. (2005). Strategies for fabricating atom probe specimens with a dual beam FIB. *Ultramicroscopy* **102**, 287–298. http://www.scopus.com/inward/record.url?eid=2-s2.0-13444273231&partnerID=tZOtx3y1.

MONTOYA, E., BALS, S., ROSSELL, M. D., SCHRYVERS, D. & VAN TENDELOO, G. (2007). Evaluation of top, angle, and side cleaned FIB samples for TEM analysis. *Microscopy Research and Technique* **70**, 1060–1071.

MOUTON, I., CHANG, Y., CHAKRABORTY, P., WANG, S., STEPHENSON, L. T., BEN BRITTON, T. & GAULT, B. (2021). Hydride growth mechanism in zircaloy-4: Investigation of the partitioning of alloying elements. *Materialia* **15**, 101006.

MOY, C. K. S., RANZI, G., PETERSEN, T. C. & RINGER, S. P. (2011). Macroscopic electrical field distribution and field-induced surface stresses of needle-shaped field emitters. *Ultramicroscopy* **111**, 397–404.

Nobel Prize ®and the Nobel Prize ®medal design mark are registered trademarks of the Nobel Foundation Scientific Background on the Nobel Prize in Chemistry 2017 THE DEVELOPMENT OF CRYO-ELECTRON MICROSCOPY (2017). https://www.nobelprize.org/nobel_prizes/chemistry/laureates/2017/advanced-chemistryprize2017.pdf.

ORÚS, P., SIGLOCH, F., SANGIAO, S. & DE TERESA, J. M. (2021). Cryo-Focused Ion Beam-Induced Deposition of Tungsten–Carbon Nanostructures Using a Thermoelectric Plate. *Applied Sciences 2021, Vol. 11, Page 10123* **11**, 10123.

PARK, Y. K., NAGAI, T., TAKAI, M., LEHRER, C., FREY, L. & RYSSEL, H. (1999). Microanalysis of FIB induced deposited Pt films using ion microprobe. *Proceedings of the International Conference on Ion Implantation Technology* **2**, 1137–1139.

PARMENTER, C. D. & NIZAMUDEEN, Z. A. (2020). Cryo-FIB-lift-out: practically impossible to practical reality. *Journal of Microscopy* jmi.12953. https://onlinelibrary.wiley.com/doi/abs/10.1111/jmi.12953 (Accessed September 14, 2020).

PEREA, D. E., GERSTL, S. S. A., CHIN, J., HIRSCHI, B. & EVANS, JAMES. E. (2017). An environmental transfer hub for multimodal atom probe tomography. *Advanced structural and chemical imaging* **3**, 12.

PROSA, T. J. & LARSON, D. J. (2017). Modern Focused-Ion-Beam-Based Site-Specific Specimen Preparation for Atom Probe Tomography. *Microscopy and Microanalysis* **23**, 194–209.

REVEALED, I., YUWONO, V. M., BURROWS, N. D., SOLTIS, J. A. & LEE PENN, R. (2010). Oriented aggregation: Formation and transformation of mesocrystal. *Journal of the American Chemical Society* **132**, 2163–2165. https://pubs.acs.org/doi/full/10.1021/ja909769a (Accessed November 10, 2022).

RINGER, S. (2006). Recent Advances in Atom Probe Tomography and Applications to Understanding Nanomaterials. *Microscopy and Microanalysis* **12**, 220–221.

SALVADOR-PORROCHE, A., SANGIAO, S., PHILIPP, P., CEA, P. & TERESA, J. M. DE (2020). Optimization of Pt-C Deposits by Cryo-FIBID: Substantial Growth Rate Increase and Quasi-Metallic Behaviour. *Nanomaterials 2020, Vol. 10, Page 1906* **10**, 1906. https://www.mdpi.com/2079-4991/10/10/1906/htm (Accessed September 13, 2021).

SASAKI, T. T., SEPEHRI-AMIN, H., UZUHASHI, J., OHKUBO, T. & HONO, K. (2022). Complementary and correlative (S)TEM/APT analysis of functional and structural alloys. *MRS Bulletin* **47**, 688–695. https://link.springer.com/article/10.1557/s43577-022-00374-7 (Accessed November 10, 2022).

SAXEY, D. W., MOSER, D. E., PIAZOLO, S., REDDY, S. M. & VALLEY, J. W. (2018). Atomic worlds: Current state and future of atom probe tomography in geoscience. *Scripta Materialia* **148**, 115–121.

SCHREIBER, D. K., PEREA, D. E., RYAN, J. V, EVANS, J. E. & VIENNA, J. D. (2018). A method for site-specific and cryogenic specimen fabrication of liquid/solid interfaces for atom probe tomography. *Ultramicroscopy* **194**, 89–99.

SEOL, J. B., KWAK, C. M., KIM, Y. T. & PARK, C. G. (2016). Understanding of the field evaporation of surface modified oxide materials through transmission electron microscopy and atom probe tomography. *Applied Surface Science* **368**, 368–377.

SPURGEON, S. R., SUSHKO, P. V., DEVARAJ, A., DU, Y., DROUBAY, T. & CHAMBERS, S. A. (2018). Onset of phase separation in the double perovskite oxide La2MnNiO6. *Physical Review B* **97**, 134110. https://journals.aps.org/prb/abstract/10.1103/PhysRevB.97.134110 (Accessed November 10, 2022).

TAYLOR, K. A. & GLAESER, R. M. (1974). Electron Diffraction of Frozen, Hydrated Protein Crystals. *Science* **186**, 1036–1037. https://www.science.org/doi/10.1126/science.186.4168.1036 (Accessed November 10, 2022).

———— (1976). Electron microscopy of frozen hydrated biological specimens. *Journal of Ultrastructure Research* **55**, 448–456.

THOMPSON, K., LAWRENCE, D., LARSON, D. J., OLSON, J. D., KELLY, T. F. & GORMAN, B. (2007). In situ site-specific specimen preparation for atom probe tomography. *Ultramicroscopy* **107**, 131–139.

WILKES, T. J., TITCHMAR, J. M., SMITH, G. D. W., SMITH, D. A., MORRIS, R. F., JOHNSTON, S., GODFREY, T. J. & BIRDSEYE, P. (1972). Fracture of Field-Ion Microscope Specimens. *Journal of Physics D-Applied Physics* **5**, 2226–2230.

WINTER, D. A. M. DE & MULDERS, J. J. L. (2007). Redeposition characteristics of focused ion beam milling for nanofabrication. *Journal of Vacuum Science & Technology B: Microelectronics and Nanometer Structures Processing, Measurement, and Phenomena* **25**, 2215.

Zachman, M. J., Tu, Z., Choudhury, S., Archer, L. A. & Kourkoutis, L. F. (2018). Cryo-STEM mapping of solid–liquid interfaces and dendrites in lithium-metal batteries. *Nature* **560**, 345–349. https://doi.org/10.1038/s41586-018-0397-3 (Accessed September 28, 2020).

Zhong, X. L., Haigh, S. J., Zhou, X. & Withers, P. J. (2020). An in-situ method for protecting internal cracks/pores from ion beam damage and reducing curtaining for TEM sample preparation using FIB. *Ultramicroscopy* **219**, 113135.

Zhu, G., Sushko, M. L., Loring, J. S., Legg, B. A., Song, M., Soltis, J. A., Huang, X., Rosso, K. M. & de Yoreo, J. J. (2021). Self-similar mesocrystals form via interface-driven nucleation and assembly. *Nature 2021 590:7846* **590**, 416–422. https://www.nature.com/articles/s41586-021-03300-0 (Accessed November 10, 2022).